
\magnification=1200
\def\ni{\noindent}
\def\.{\mathaccent 95}

\def\ga{\gamma}

\def\De{\Delta}

\def\frac#1#2{{\textstyle{{#1}\over {#2}}}}
\def\ni{\noindent}
\def\lsim{\mathrel{\rlap{\lower4pt\hbox{\hskip1pt$\sim$}}
    \raise1pt\hbox{$<$}}}
\def\gsim{\mathrel{\rlap{\lower4pt\hbox{\hskip1pt$\sim$}}
    \raise1pt\hbox{$>$}}}
\def\sqr#1#2{{\vcenter{\vbox{\hrule height.#2pt
         \hbox{\vrule width.#2pt height#1pt \kern#1pt
         \vrule width.#2pt}
         \hrule height.#2pt}}}}

\newbox\grsign \setbox\grsign=\hbox{$>$} \newdimen\grdimen \grdimen=\ht\grsign
\newbox\simlessbox \newbox\simgreatbox
\setbox\simgreatbox=\hbox{\raise.5ex\hbox{$>$}\llap
     {\lower.5ex\hbox{$\sim$}}}\ht1=\grdimen\dp1=0pt
\setbox\simlessbox=\hbox{\raise.5ex\hbox{$<$}\llap
     {\lower.5ex\hbox{$\sim$}}}\ht2=\grdimen\dp2=0pt

%
%

\def\ref#1  {\noindent \hangindent=24.0pt \hangafter=1 {#1} \par}

\def\doublespace {\smallskipamount=6pt plus2pt minus2pt
                  \medskipamount=12pt plus4pt minus4pt
                  \bigskipamount=24pt plus8pt minus8pt
                  \normalbaselineskip=24pt plus0pt minus0pt
                  \normallineskip=2pt
                  \normallineskiplimit=0pt
                  \jot=6pt
                  {\def\smallskip {\vskip\smallskipamount}}
                  {\def\medskip   {\vskip\medskipamount}}
                  {\def\bigskip   {\vskip\bigskipamount}}
                  {\setbox\strutbox=\hbox{\vrule 
                    height17.0pt depth7.0pt width 0pt}}
                  \parskip 12.0pt
                  \normalbaselines}


\centerline{\bf Importance of an Astrophysical Perspective for
Textbook Relativity}
\centerline{Eric G. Blackman}
\centerline{Institute of Astronomy, Madingley Road, Cambridge, CB3 OHA}
\medskip
\centerline{(in press, European Journal of Physics)}
\medskip
\doublespace
\medskip

\centerline{\bf ABSTRACT}

The importance of a teaching a clear definition of the ``observer'' 
in special relativity is highlighted using a simple astrophysical example
from the exciting current research area of ``Gamma-Ray Burst'' astrophysics.
The example shows that a source moving relativistically toward a single
observer at rest exhibits a time ``contraction'' rather than a 
``dilation'' because the light travel time between the source
and observer decreases with time.  
Astrophysical applications of special relativity 
complement idealized examples with real applications and very 
effectively exemplify the role of a finite light travel time.  
 
\vfill
\eject

The basics of special relativity such as length contraction, 
time dilation, and simultaneity 
are commonly taught using idealized examples of
moving trains, rocketed twins, and cars moving though
small garages with doors that open and close at either end. 
However, very important applications of special relativity 
to the real world come from astrophysics, where many sources
with relativistic outflows are seen and
the observed effects of special relativity in action 
are ubiquitous.  

An important difference between idealized examples
and astronomical examples is in the definition of the ``observer.''
In astronomy, the observer is always a single detector at a fixed 
location in space.  The difference in light travel times 
to the detector between successive observations of a relativistically
moving source is thus fundamental to every observation.
In idealized examples however, it is possible to imagine that every space
time point has a measuring rod and clock.
Because astrophysical examples highlight the consequences
of the above difference, and generally show very real effects of special 
relativity in action, classroom teaching benefits from
an astrophysical perspective.  Below I derive an example of this
from the current frontier field of Gamma-Ray Burst astronomy.
The example emphasizes the importance of a clear definition of observer,
and how a source with a time-varying luminosity
moving relativistically toward a single detector would
be seen by that detector to exhibit an effective time-contraction(!) rather
than a time dilation. This is because the light travel
distance to the detector is constantly decreasing during the motion.

First a little background:
Gamma-Ray Bursts (GRBs) are notoriously mysterious astrophysical
sources which emit bursts of gamma-ray emission lasting
$\sim 0.1$ to $\sim 100$ sec without repeating.$^{1}$ 
A new one appears every day
and the population is isotropically distributed on the sky.
This isotropy has been one key feature suggesting that they are located 
at cosmological distances.  This has now been 
confirmed directly for at least 1 burst$^{2}$. 
Each burst requires 
a luminosity of order $\gsim 10^{18}$ times that of the Sun! 
Though little is known about the actual source
engines, relativistic flows are believed to be important
to GRB since they significantly relax the model constraints of
non-relativistic models.$^{1}$    
In short, bulk flow Lorentz factors of  $\ga >100$ are likely required,
the highest of any source in nature.

The study of variability times in these relativistic GRB highlights the
importance of an astrophysical perspective in clearly defining
an ``observer'' in special relativity.  Consider one such variability time, 
taken to be the duration of an individual flare within a spiky GRB light curve.
Imagine the source to be a highly relativistic expanding shell 
with constant Lorentz factor $\ga$ emitting photons toward a distant detector at rest with respect to the shell. For $\ga >>1$, the only emission 
seen at the detector is produced by the shell material moving
within an angle $\theta \sim 1/\gamma$ to the detector line of sight.
Suppose that in the co-moving frame, the time duration of
a single flare emitted by material moving directly along the line
of sight is given by $\De t_C \equiv t_{2C}-t_{1C}$.
Assume that the flare remains at the same location along the line of
sight in the co-moving 
frame, in other words let $x_{1C}=x_{2C}$, where the variable $x$ measures
displacement along the line of sight and $C$ indicates the co-moving
frame.  The duration as measured at the detector is then given by 
$$\De t_D= \ga\De t_{C}-\De x_D/c, \eqno(1)$$
where 
$\De x_D\equiv x_{1D}-x_{2D}$, and  
$x_{1D}$ and $x_{2D}$ are the shell positions measured in the
detector frame at the beginning and end of the flare respectively.  
Now, note that  $\De x_D = \ga v\De t_C$, so that
$$\De t_D= \ga \De t_C(1-v/c)\sim \De t_C / 2\gamma 
\sim \De x_D/(2c\ga^2),  \eqno(2)$$
where the last similarity uses $v\sim c$.
Thus, note that $\gamma>>1$ means the variability time
measured at the detector, 
$\De t_D$, is short compared to the co-moving $\De t_C$.

 

Let us reflect on this result.  
The derivation of (2) highlights  why the statement commonly
made in special relativity that ``a clock moving relative to an
observer runs slow'' $^{3}$ must 
be interpreted very carefully.  Strictly speaking, this statement is  
actually $opposite$ to (2) if the observer is the single detector.
That a single detector does not generally measure the standard time 
dilation results from the fact that the photon travel distances to the 
detector are different from different points of the motion. This accounts 
for the presence of the last term of (1), and the appearance of $\ga$ in the 
denominator rather than in the numerator on the right side of (2).
  
Note that while many textbook treatments $^{4,5,6}$ are 
careful in defining what is an appropriate observer to measure
time dilation and length contraction
(ie. appropriate observer = a system of rods and clocks used throughout the frame which
can measure at the position of each event), some textbooks are not.
The interpretation of detected variability in ultra-relativistic
GRB shows through (2), why a clear definition of the observer
is absolutely essential.  An observer in astrophysics means a single
detector, but as shown above, such an observer would not generally  
measure the standard time dilation and length contraction from a moving source unless the distance between the source and observer at each measured 
space-time event were the same.

Some effects of the different 
photon travel distances from a relativistically expanding source shell 
toward a single detector were 
considered in the classic paper by Rees$^{7}$, where 
apparent super-luminal motion in astrophysical sources
was first predicted.  Recent excitement
and attention toward GRB and relativistic flows 
provides the opportunity to highlight why
an astrophysical perspective can be extremely important to
the study of fundamental special relativity and can help bring
the taste of a current research topic into the classroom.

\noindent Thanks to L. Rodriguez-Williams and A. Fabian for discussions.

\ni 1. M.J. Rees, preprint, http://xxx.lanl.gov/astro-ph/9701162, (1997).

\ni 2. M.R. Metzger  et al., {\it Nature}, {\bf 387}, 878 (1997)







\ni 3. M.H. Nayfeh \& M.K. Brussel, {\it Electricity and Magnetism}, 
(Wiley, New York, 1985), p564.

\ni 4. E.M. Purcell,  {\it Electricity and Magnetism}, (McGraw-Hill, New York, 1985), p452.

\ni 5. E.F. Taylor, \& J.A. Wheeler, {\it Spacetime Physics}, 
(Freeman, San Francisco, 1966) p18.

\ni 6. W. Rindler,  {\it Essential Relativity}, (Spinger-Verlag,  New York, 1977), p. 30.

\ni 7. M.J. Rees, {\it Nature}, {\bf 211}, 468, (1966).




\end